\documentclass[reprint,prl,amsmath,amssymb,superscriptaddress,noffotinbib,preprintnumbers]{revtex4-1}
\usepackage{url}
\usepackage{color}
\usepackage{isotope}
\usepackage{pstricks}
\usepackage{colordvi}
\usepackage{epsfig}% Include figure files
\usepackage{graphicx}% Include figure files
\usepackage{dcolumn}% Align table columns on decimal point
\usepackage{bm}% bold math
\usepackage{amsmath}
\usepackage{amssymb}
\usepackage{hyperref}
\usepackage{slashed}
\usepackage{lineno}
\usepackage{isotope}

\hypersetup{
    colorlinks=true, %set true if you want colored links
    linkcolor=blue,  %choose some color if you want links to stand out
   citecolor=blue
}

\begin{document}

%-----------------------------------------------------------------------------------------------------------------------------------------------------------------------%

\title{First Measurement of the {Ar}$(e,e^\prime)X$ Cross Section at Jefferson Laboratory}

\author{H.~Dai} \affiliation{Center for Neutrino Physics, Virginia Tech, Blacksburg, Virginia 24061, USA}
\author{M.~Murphy} \affiliation{Center for Neutrino Physics, Virginia Tech, Blacksburg, Virginia 24061, USA}
\author{V.~Pandey} 
\email{vishvas.pandey@vt.edu}
\affiliation{Center for Neutrino Physics, Virginia Tech, Blacksburg, Virginia 24061, USA}
\author{D.~Abrams} \affiliation{Department of Physics, University of Virginia, Charlottesville, Virginia 22904, USA}
\author{D.~Nguyen} \affiliation{Department of Physics, University of Virginia, Charlottesville, Virginia 22904, USA}
\author{B.~Aljawrneh} \affiliation{North Carolina Agricultural and Technical State University, Greensboro, North Carolina 27401, USA}
\author{S.~Alsalmi} \affiliation{Kent State University, Kent, Ohio 44242, USA}
\author{A.~M.~Ankowski} \affiliation{SLAC National Accelerator Laboratory, Stanford University, Menlo Park, California 94025, USA}
\author{J.~Bane} \affiliation{The University of Tennessee, Knoxville, Tennessee 37996, USA}
\author{S.~Barcus} \affiliation{The College of William and Mary, Williamsburg, Virginia 23187, USA}
\author{O.~Benhar} \affiliation{INFN and Dipartimento di Fisica, Sapienza Universit\`{a} di Roma, I-00185 Roma, Italy}
\author{V.~Bellini} \affiliation{INFN, Sezione di Catania, Catania, 95123, Italy}
\author{J.~Bericic} \affiliation{Thomas Jefferson National Accelerator Facility, Newport News, Virginia 23606, USA}
\author{D.~Biswas} \affiliation{Hampton University, Hampton, Virginia 23669, USA}
\author{A.~Camsonne} \affiliation{Thomas Jefferson National Accelerator Facility, Newport News, Virginia 23606, USA}
\author{J.~Castellanos} \affiliation{Florida International University, Miami, Florida 33181, USA}
\author{J.-P.~Chen} \affiliation{Thomas Jefferson National Accelerator Facility, Newport News, Virginia 23606, USA}
\author{M.~E.~Christy} \affiliation{Hampton University, Hampton, Virginia 23669, USA}
\author{K.~Craycraft} \affiliation{The University of Tennessee, Knoxville, Tennessee 37996, USA}
\author{R.~Cruz-Torres} \affiliation{Massachusetts Institute of Technology, Cambridge, Massachusetts 02139, USA}
\author{D.~Day} \affiliation{Department of Physics, University of Virginia, Charlottesville, Virginia 22904, USA}
\author{S.-C.~Dusa} \affiliation{Thomas Jefferson National Accelerator Facility, Newport News, Virginia 23606, USA}
\author{E.~Fuchey} \affiliation{University of Connecticut, Storrs, Connecticut 06269, USA}
\author{T.~Gautam} \affiliation{Hampton University, Hampton, Virginia 23669, USA}
\author{C.~Giusti} \affiliation{Dipartimento di Fisica, Universit\`{a} degli Studi di Pavia and INFN, Sezione di Pavia,  I-27100 Pavia, Italy}
\author{J.~Gomez} \affiliation{Thomas Jefferson National Accelerator Facility, Newport News, Virginia 23606, USA}
\author{C.~Gu} \affiliation{Duke University, Durham, North Carolina 27708, USA}
\author{T.~Hague} \affiliation{Kent State University, Kent, Ohio 44242, USA}
\author{J.-O.~Hansen} \affiliation{Thomas Jefferson National Accelerator Facility, Newport News, Virginia 23606, USA}
\author{F.~Hauenstein} \affiliation{Old Dominion University, Norfolk, Virginia 23529, USA}
\author{D.~W.~Higinbotham} \affiliation{Thomas Jefferson National Accelerator Facility, Newport News, Virginia 23606, USA}
\author{C.~Hyde} \affiliation{Old Dominion University, Norfolk, Virginia 23529, USA}
\author{C.~M.~Jen} \affiliation{Center for Neutrino Physics, Virginia Tech, Blacksburg, Virginia 24061, USA}
\author{C.~Keppel} \affiliation{Thomas Jefferson National Accelerator Facility, Newport News, Virginia 23606, USA}
\author{S.~Li} \affiliation{University of New Hampshire, Durham, New Hampshire 03824, USA}
\author{R.~Lindgren} \affiliation{Department of Physics, University of Virginia, Charlottesville, Virginia 22904, USA}
\author{H.~Liu} \affiliation{Columbia University, New York, New York 10027, USA}
\author{C.~Mariani} \affiliation{Center for Neutrino Physics, Virginia Tech, Blacksburg, Virginia 24061, USA}
\author{R.~E.~McClellan} \affiliation{Thomas Jefferson National Accelerator Facility, Newport News, Virginia 23606, USA}
\author{D.~Meekins} \affiliation{Thomas Jefferson National Accelerator Facility, Newport News, Virginia 23606, USA}
\author{R.~Michaels} \affiliation{Thomas Jefferson National Accelerator Facility, Newport News, Virginia 23606, USA}
\author{M.~Mihovilovic} \affiliation{Jozef Stefan Institute, Ljubljana 1000, Slovenia}
\author{M.~Nycz} \affiliation{Kent State University, Kent, Ohio 44242, USA}
\author{L.~Ou} \affiliation{Massachusetts Institute of Technology, Cambridge, Massachusetts 02139, USA}
\author{B.~Pandey} \affiliation{Hampton University, Hampton, Virginia 23669, USA}
\author{K.~Park} \affiliation{Thomas Jefferson National Accelerator Facility, Newport News, Virginia 23606, USA}
\author{G.~Perera} \affiliation{Department of Physics, University of Virginia, Charlottesville, Virginia 22904, USA}
\author{A.~J.~R.~Puckett} \affiliation{University of Connecticut, Storrs, Connecticut 06269, USA}
\author{S.~N.~Santiesteban} \affiliation{University of New Hampshire, Durham, New Hampshire 03824, USA}
\author{S.~\v{S}irca} \affiliation{University of Ljubljana, Ljubljana 1000, Slovenia} \affiliation{Jozef Stefan Institute, Ljubljana 1000, Slovenia}
\author{T.~Su} \affiliation{Kent State University, Kent, Ohio 44242, USA}
\author{L.~Tang} \affiliation{Hampton University, Hampton, Virginia 23669, USA}
\author{Y.~Tian} \affiliation{Shandong University, Shandong, 250000, China}
\author{N.~Ton} \affiliation{Department of Physics, University of Virginia, Charlottesville, Virginia 22904, USA}
\author{B.~Wojtsekhowski} \affiliation{Thomas Jefferson National Accelerator Facility, Newport News, Virginia 23606, USA}
\author{S.~Wood} \affiliation{Thomas Jefferson National Accelerator Facility, Newport News, Virginia 23606, USA}
\author{Z.~Ye} \affiliation{Physics Division, Argonne National Laboratory, Argonne, Illinois 60439, USA}
\author{J.~Zhang} \affiliation{Department of Physics, University of Virginia, Charlottesville, Virginia 22904, USA}

\collaboration{The Jefferson Lab Hall A Collaboration}

%-----------------------------------------------------------------------------------------------------------------------------------------------------------------------%

\begin{abstract}

The success of the ambitious programs of both long- and short-baseline neutrino-oscillation experiments employing liquid-argon time-projection chambers will greatly rely on the precision with which the weak response of the argon nucleus can be estimated. In the E12-14-012 experiment at Jefferson Lab Hall A, we studied the properties of the argon nucleus by scattering a high-quality electron beam off a high-pressure gaseous argon target. Here, we present the measured $^{40}$Ar$(e,e^{\prime})$ double differential cross section at incident electron energy $E=2.222$~GeV and scattering angle $\theta = 15.54^\circ$. The data cover a broad range of energy transfers, where quasielastic scattering and delta production are the dominant reaction mechanisms. The result for argon is compared to our previously reported cross sections for titanium and carbon, obtained in the same kinematical setup.
\end{abstract}

\preprint{JLAB-PHY-18-2859}
\preprint{SLAC-PUB-17338}

%-----------------------------------------------------------------------------------------------------------------------------------------------------------------------%

\maketitle

\par Precise determination of charge-parity (CP) symmetry violation in the lepton sector---necessary to shed light on the matter-antimatter asymmetry in the universe---is among the highest priorities of particle physics. Over the next two decades, this issue will be a primary science goal of the Deep Underground Neutrino Experiment (DUNE)~\cite{DUNE:2018}, together with a search for proton decay, measurement of the electron-neutrino flux from a core-collapse supernova---should one occur in our galaxy during the lifetime of DUNE---and search for physics beyond the standard model.
\par In the next few years, the Short-Baseline Neutrino (SBN) program~\cite{SBN:2015} at Fermilab will provide a definitive answer to the question of the existence of sterile neutrinos, which could be the source of electron-like events recently reported with statistical significance 4.8$\sigma$ by the MiniBooNE Collaboration~\cite{MiniBooNE:2018}.
\par Both DUNE and the SBN program (will) employ liquid-argon time-projection chambers as their detectors, the advantages of which are low threshold momenta for particle detection and high spatial resolution, allowing (among others) for precise neutrino-energy reconstruction and distinguishing photons from electrons. As a consequence, the success of both programs in studying neutrino oscillations with unprecedented precision will greatly rely on the precision with which we understand the complexity of nuclear effects in argon and the precision with which we are able to estimate its response to electroweak probes.
\par It is important to realize that, although the near detector facilities of DUNE will play a fundamental role in the reduction of systematic uncertainties, alone they will not be sufficient to determine the cross sections with the precision necessary to achieve the objectives of DUNE~\cite{Acciarri:2015uup}. At beam energies in the few-GeV region, the observed event kinematics cannot be readily translated to the true value of neutrino energy, owing to detector effects, and the procedure of energy reconstruction heavily relies on the nuclear model used in Monte Carlo (MC) simulations~\cite{Ankowski:2015kya}. Even for functionally identical near and far detectors, the spectrum reconstructed in the near detector is very different from the one in the far detector. This is a consequence not only of neutrino oscillations, but also of differences in particle containment and angular acceptance, and of the strong angular dependence of the flux, which makes important the difference between the solid angle probed by near and far detectors, even in the absence of the oscillations. As the CP-violation sensitivity of DUNE critically depends on systematic uncertainties, even their modest reduction has a meaningful impact on the running time necessary to achieve the physics objectives.
\par In the ongoing oscillation experiments~\cite{Abe:2017uxa,NOvA:2018gge}, the uncertainties related to nuclear effects in neutrino-nucleus interactions have become one of the major sources of systematics~\cite{Benhar:2015, Alvarez-Ruso:2017}, despite extensive use of near-detector data to constrain the nuclear models employed in MC simulations. As different probe's energies and reaction mechanisms are intertwined in neutrino-scattering data, it is difficult to identify, diagnose, and remedy potential shortcomings of nuclear models. However, electron-scattering measurements with targets and kinematics of interest to neutrino experiments give an excellent opportunity to validate and improve the description of nuclear effects~\cite{Ankowski:2016jdd}. Considering that there is a large body of electron-scattering data available for carbon (and limited availability of data for oxygen) the situation for argon is woefully inadequate, with only one dataset currently available: the inclusive electron-scattering spectrum measured at Frascati National Laboratory (LNF) using the electron-positron collider ADONE and a jet target at incident electron energy E$=$700~MeV and scattering angle $\theta = 32^\circ$~\cite{Anghinolfi:1995}. Argon can be expected to be more challenging to describe than oxygen and carbon, as a significantly heavier nucleus that is additionally isospin asymmetric. This asymmetry is of fundamental importance for the CP-violation measurement in DUNE, to be based on analysis of the difference between the neutrino and antineutrino event distributions. Availability of a new precise dataset for electron scattering off argon is therefore vital, in order to provide a testbed and stimulate further development of theoretical models of nuclear response to electroweak interactions~\cite{Ankowski:2015, Rocco:2016, Vagnoni:2017, Lovato:2016, Meucci:2014, Lalakulich:2012, Nieves:2012, Martini:2010, Pandey:2014, Megias:2016} in the kinematic region of interest to neutrino experiments.
\par To address this issue, we performed a dedicated experiment at Jefferson Lab (JLab) to study electron scattering from argon and titanium nuclei~\cite{Proposal:2014}. The experiment, E12-14-012, collected high statistics data in JLab Hall A during February and March 2017. We recently reported {Ti}$(e,e^\prime){X}$ and {C}$(e,e^\prime){X}$ cross section results~\cite{Dai:2018}. Here, we present the first argon results of the experiment, {Ar}$(e,e^\prime){X}$ cross section at beam energy $E = 2.222$~GeV and electron scattering angle $\theta = 15.54^\circ$, and its comparison with our previously reported cross sections for the titanium and carbon nuclei in the same kinematics~\cite{Dai:2018}.
\par In the analyzed $(e,e^\prime)$ process, $ e + A \rightarrow e^\prime  + X $, an electron of four-momentum $k \equiv (E, {\bf k})$ scatters off a nuclear target $A$. The energy and scattering angle of the outgoing electron of four-momentum $k^\prime\equiv (E^\prime, {\bf k}^\prime )$ are measured while the hadronic final state remains undetected. The squared four-momentum transfer in the process is $q^2=-Q^2$, with $q = k - k^\prime \equiv (\omega, {\bf q})$.
\par A continuous-wave electron beam of energy E$=$2.222~GeV (known with accuracy better than 0.1\%) was supplied by the Continuous Electron Beam Accelerator Facility (CEBAF) at JLab. The current and position of the beam, the latter being critical for vertex reconstruction and momentum calculation of scattered electrons, were monitored by resonant radio-frequency cavities (beam current monitors or BCMs) and cavities with four antennas (beam position monitors or BPMs), respectively. Harp scanners, which moved a thin wire through the beam, were used to measure its size. To eliminate the possibility of overheating the target by the deposited beam energy, the beam was rastered with a 2 $\times$ 2~mm$^2$ raster system, to increase the effective spot size and reduce the energy density.
\par The gaseous argon target, with a thickness of 1.455$\pm$0.005~g/cm$^2$, was contained in a 25~cm long cell with thin aluminum entry and exit windows of respectively 0.25 and 0.28~mm thickness. To account for the background contribution from electrons scattered from the wall of the argon target cell measurements were also performed on a dummy target, aluminum foils mounted on separate frames located at positions corresponding to the entry and exit windows of the cell. The thickness of the entry and exit aluminum foils was 0.889$\pm$0.002~g/cm$^2$ and matched the radiation length of the argon target.
\par The scattered electrons were detected in the Left High-Resolution Spectrometer (LHRS) positioned at $\theta = 15.54^\circ$. The LHRS was equipped with superconductive magnets and a detector package for tracking, timing and particle identification~\cite{Alcorn:2004, footnote}. The scattered electrons first passed through three superconducting quadrupole magnets (Q) and one dipole magnet (D) arranged in QQDQ configuration. This arrangement provided a large acceptance in both angle and momentum, and good resolution in momentum ($\sim$10$^{-4}$), position ($\sim$10$^{-3}$~m), and angle ($\sim$1.0~mrad). The electrons then entered the detector package consisting of vertical drift chambers (VDCs), threshold \v{C}erenkov counter, scintillator detectors and a lead-glass calorimeter. The data-acquisition (DAQ) electronics was triggered when an electron passes through two scintillator detectors planes (with a logical {\textsc{and}}) and simultaneously produced a signal in the gas CO$_2$ \v{C}erenkov counter, mounted between the two scintillator planes. Electron-pion separation was achieved with the combined amplitude response of the gas \v{C}erenkov and Pb-glass shower counters. The tracking information (position and direction) was reconstructed in the VDCs utilizing a reconstruction matrix obtained from special optics-calibration runs.
\par The electron yield ($Y$) for $i$th bin in scattered electron energy ($E^\prime$) is obtained as
\begin{equation}
 Y^{i}= (N_S^i \times {{DAQ}}_{\text{pre-scale}})/(LT \times \epsilon).
\end{equation}
\begin{figure}[t!]
\centering
\includegraphics[width=1.0\columnwidth]{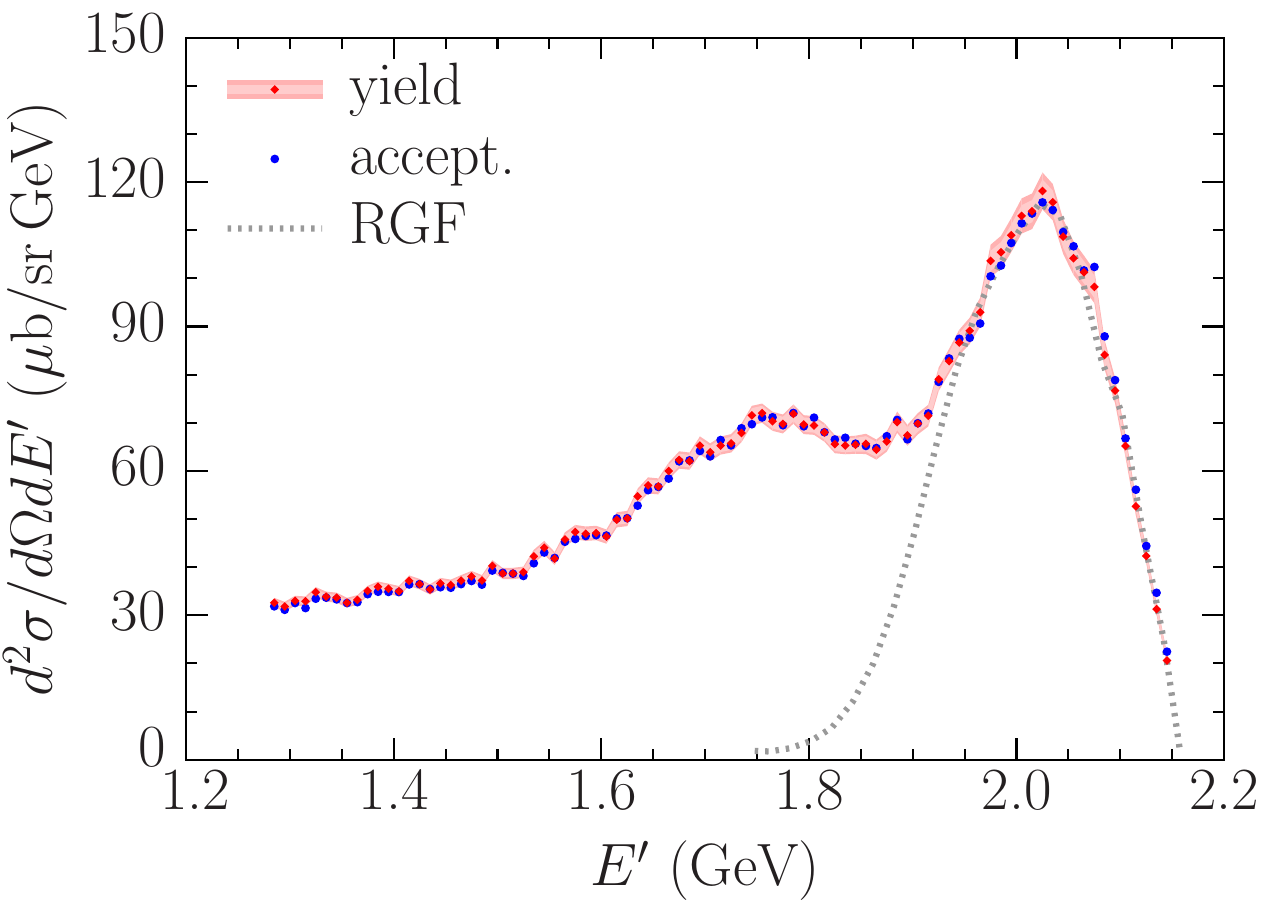}
\caption{(color online). Double differential cross section for the {Ar}$(e,e^\prime)$ process, extracted with two different methods, at beam energy of 2.222~GeV and scattering angle of 15.54$^\circ$. The inner and outer bars correspond to the statistical and total uncertainty, respectively. The dotted curve represents the quasielastic calculations obtained within the RGF formalism described in Ref.~\cite{Meucci:2003}.}
\label{xsec_Ar}
\end{figure}
Here, $N_S^i$ is the number of scattered electrons recorded, $LT$ is the live-time fraction, and $\epsilon$ is the total detection efficiency. The hardware trigger is configured to accept only every~$n = DAQ_\text{pre-scale}$ raw triggers. The $\sim$10$\mu$A beam rastered over 2$\times$2~mm$^2$ deposits enough energy into the target that its density change must be taken into consideration when extracting the cross section. This is done through a {\it target-boiling effect} study in which the beam current is ramped in steps from zero current to $\sim$20 $\mu$A and the scattering yield determined~\cite{Target}. From this a correction to the zero current density can be made and applied to all the runs. The yield is also corrected for the background ($\sim$0.2\%) remaining after the dummy cell is subtracted. Once the yield is determined, the cross section can be extracted either by the {\it {acceptance-correction method}} or by the {\it {yield-ratio method}}.
\par In the acceptance-correction method, for each bin in $\Delta E \Delta\Omega$, the cross section is obtained as
\begin{equation}
 d^{2}\sigma/d\Omega dE' =  Y(E',\theta) / [(\Delta E \Delta\Omega) A(E',\theta) L].
\end{equation}
where, $Y(E',\theta)$ and $A(E',\theta)$ are yield and acceptance for a given bin, respectively, and $L$ is the integrated luminosity obtained using MC and validated with the solid Al target (dummy cell) and C foils (the optics target). In the yield-ratio method, the cross section for each bin is computed as the product of the MC cross section~\cite{Arrington:1999} times the ratio of the data to simulation yields
\begin{equation}
 d^{2}\sigma/d\Omega dE' = (d^{2}\sigma/d\Omega dE')_{\text{MC}} \times [Y(E',\theta)/ Y_{\text{MC}}(E',\theta)].
\end{equation}
The MC cross section is a fit to the existing data including preliminary Hall C data~\cite{HallC}. The MC includes the radiative corrections computed using the peaking approximation~\cite{Tsai:1969} and Coulomb corrections implemented with an effective momentum approximation~\cite{Aste:2005}, further accounting for the change in radiation length of the target due to the target-boiling effect.
\par Figure~\ref{xsec_Ar} shows the measured Ar$(e,e^\prime )$ double differential cross section as a function of the energy of the scattered electron, $E^\prime$, extracted with the yield-ratio and the acceptance-correction methods. Both methods yield the cross-section results in very good agreement, with marginal differences observed only in the region of $E^{\prime}$ above the quasielastic peak (i.e. $\omega$ below the peak), where the event statistics are limited and the systematic uncertainties of the acceptance method are larger. The primary difference between the two methods is the fact that the yield-ratio method relies more on the predictions of the cross section model in the MC but the agreement of the two methods strengthens our confidence in both procedures. The measured cross section covers a broad range of scattered electron energy ranging from $\sim$1.3 to $\sim$2.2~GeV. The kinematical coverage includes both the quasielastic and delta-production peaks, and further extends to the deep-inelastic scattering region. The total uncertainties remain below $\sim$4.0\% corresponding to the statistical (1.7--2.9\%) and the systematic (1.8--3.0\%) uncertainties summed in quadrature. A detailed list of the uncertainties is given in Table~\ref{tab:syst}.
\begin{table}[h!]
\caption{Uncertainties associated with the presented Ar$(e,e^\prime )$ cross section. Numbers represent upper limits or the range for the uncertainties that vary between different kinematical regions.}
\label{tab:syst}
\begin{center}
\begin{tabular}{l c c c}
\\
\hline\hline
{1. Total statistical uncertainty} &						& 1.7--2.9\% &\\
{2. Total systematic uncertainty} &						& 1.8--3.0\% &\\
\phantom{1. }a.~Beam charge and beam energy & 			& 0.3\% &\\
\phantom{1. }b.~Beam offset $x$ and $y$ & 					& 0.4--1.0\% &\\
\phantom{1. }c.~Target thickness and boiling effect &		& 0.7\% &\\
\phantom{1. }d.~HRS offset $x$ and $y$~$+$~optics & 		& 0.6--1.2\% &\\
\phantom{1. }e.~Acceptance cut~($\theta$,$\phi$,$dp/p$) & 	& 0.6--2.4\% &\\
\phantom{1. }f.~Calorimeter and \v{C}erenkov cuts & 		        & 0.01--0.03\% &\\
\phantom{1. }g.~Cross section model &					& 1.3\%&\\
\phantom{1. }h.~Radiative and Coulomb corrections & 		        & 1.0\% &\\
\hline
\hline\\[-20pt]
\end{tabular}
\end{center}
\end{table}
\par The dotted curve of Fig.~\ref{xsec_Ar} represents the theoretical results obtained from the relativistic Green's function (RFG) approach described in Ref.~\cite{Meucci:2003}. In the RGF formalism, following assumptions based on the impulse approximation, the components of the nuclear response are written in terms of the single-particle optical-model Green's function. Final-state interactions are accounted for, consistent with the approach used in the exclusive $(e,e^{\prime}p)$ reaction, by the same complex optical potential but the formalism translates the flux lost towards inelastic channels, represented by the imaginary part of the optical potential, into the strength observed in inclusive reactions. It is apparent that this procedure leads to a remarkably good description of both shape and normalization of the data in the the quasielastic region. However, it does not include two-body currents and delta-excitation mechanisms which are clearly visible in the region of lower $E^\prime$ values (i.e. larger energy transfers).
\begin{figure}
\centering
\includegraphics[width=1.0\columnwidth]{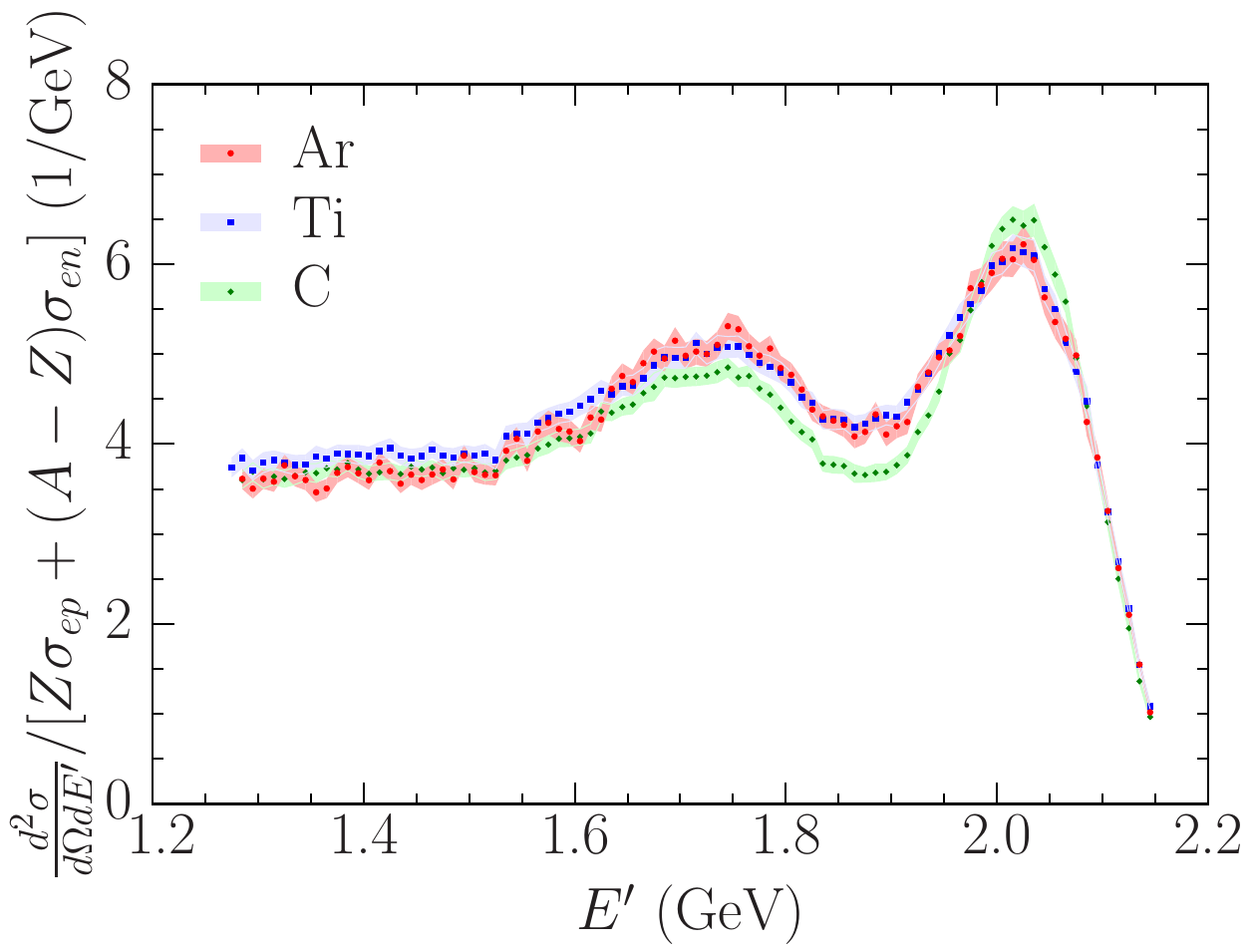}
\caption{(color online). Comparison of Ar$(e,e^{\prime})$ cross section of Fig.~\ref{xsec_Ar}, and Ti$(e,e^{\prime})$ and C$(e,e^{\prime})$ cross sections of Ref.~\cite{Dai:2018}, all in the same kinematics, presented in terms of the ratio defined by Eq.\eqref{ratio}.}
\label{fig:compare}
\end{figure}
\par In Fig.~\ref{fig:compare}, we compare the argon data to the titanium and carbon data of Ref.~\cite{Dai:2018}, taken in the same kinematical setup, corresponding to incident electron energy 2.222~GeV and scattering angle of 15.54$^\circ$. The comparison is performed in terms of the ratio defined as
\begin{align}
( d^2\sigma/d\Omega d E^\prime ) / [Z\sigma_{ep} + (A-Z)\sigma_{en}] \ ,
\label{ratio}
\end{align}
where $A$ and $Z$ are the nuclear mass number and charge, respectively, while $\sigma_{ep}$ and $\sigma_{en}$ denote the elastic electron-proton and electron-neutron cross sections stripped of the energy-conserving delta function~\cite{deForest:1983}. The results of Fig.~\ref{fig:compare}, showing that the ratios of Eq.\eqref{ratio} corresponding to argon and titanium are nearly identical to one another, appear to support the strategy underlying our experiment, aimed at exploiting titanium data to extract complementary information on nuclear effects in argon. However, the differences between the results for argon and carbon indicate significant differences in the ground-state properties of these nuclei, which are relevant in the context of MC simulations for DUNE.
\begin{figure}[h!]
\centering
\includegraphics[width=1.0\columnwidth]{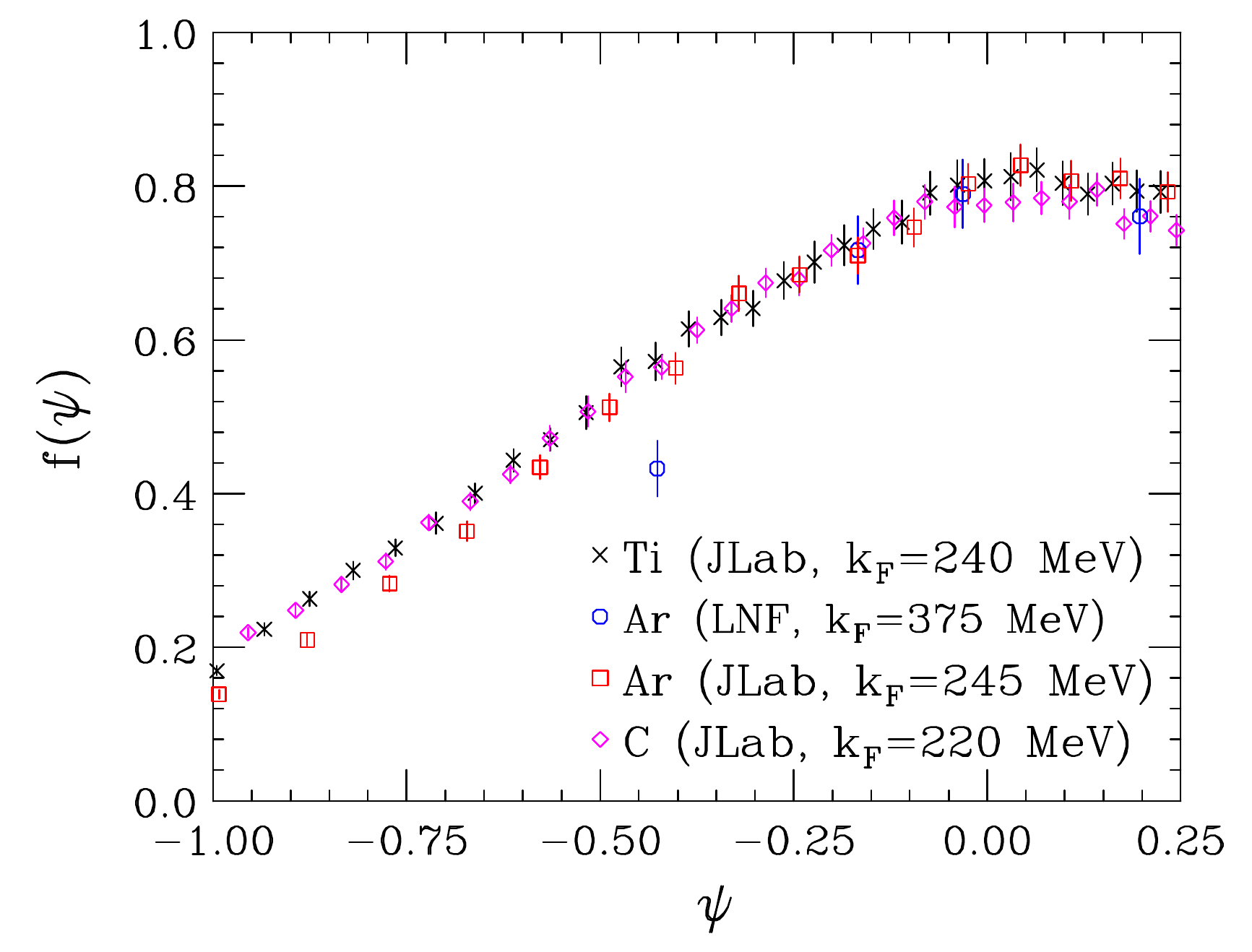}
\caption{(color online). Comparison between the scaling function of the second kind, $f(\psi)$, obtained from E12-14-012 data on Ar, Ti, and C. The $k_F$ of C is fixed to the value obtained by Moniz {\it et al.}~\cite{Moniz:1971} while the data analysis of Ti and Ar sets $k_F$ at 240 and 245~MeV, respectively. The circles are the Ar data from LNF~\cite{Anghinolfi:1995}, which turn out to prefer an inconsistently higher value of $k_F$.}
\label{fig:superscaling}
\end{figure}
%
%%%%%%%%%%%%%%%%%%%%%%%%%%%%%%%%%%%%%%%%%%%%%%%%%%%%%%%%%%%%%%%%%%%%%%%%
\par
Inclusive data corresponding to different kinematics and different targets are best compared in terms of the scaling functions of
the first and second kinds, discussed in Refs.~\cite{Sick:1980} and  \cite{Donnelly:1999}, respectively. Scaling of the first kind, or $y$-scaling,
is observed in the regime in which quasielastic single-nucleon knockout is the dominant reaction mechanism, and the effect of final state
interactions between the struck nucleon and the spectator system is negligible. The resulting scaling function, $F(y)$, 
 is determined by the target spectral function, and turns out to be largely independent of kinematics.  Scaling of the second kind, however, allows to compare data sets corresponding to different targets. The definitions of both the scaling variable $\psi$ and the scaling function 
$f(\psi)$ involve a momentum scale, which can be loosely interpreted as a nuclear Fermi momentum, $k_F$, providing a simple parametrization of the target dependence of nuclear effects.  

\par 
In Fig.~\ref{fig:superscaling}, we show the scaling functions of the second kind, $f(\psi)$, displayed as a function of the dimensionless scaling variable $\psi$. It is apparent that setting the carbon Fermi momentum to 220~MeV\textemdash the value resulting from the analysis of Moniz {\it et al.}~\cite{Moniz:1971}\textemdash the scaling of titanium and argon data is observed for $k_F =$ 240 and 245~MeV, respectively. Hence, the scaling analysis confirms the picture emerging from Fig.~\ref{fig:compare}. For comparison, we also show the scaling function $f(\psi)$ obtained using the Ar$(e,e^\prime)$ cross section at 700~MeV and 32$^\circ$, measured at the LNF electron-positron storage ring ADONE using a jet target~\cite{Anghinolfi:1995}. It turns out that the LNF data only scale at $\psi \approx 0$, and prefer a value of the Fermi momentum, $k_F$=375~MeV, much larger than that resulting from the analysis of JLab data. This inconsistency may well be the result of the normalization issue that the authors of Ref.~\cite{Anghinolfi:1995} found in their $^{16}$O cross section, as compared to the cross sections previously measured at the Bates Linear Accelerator Center~\cite{OConnell}, chosen as a reference dataset. A normalization factor of 1.19 had to be applied to the LNF $^{16}$O cross section in order to reproduce the Bates spectrum~\cite{Anghinolfi:1995}. Note that the Bates data for oxygen were obtained by subtracting cross sections corresponding  to BeO and Be targets, while the LNF experiment used a relatively {\it pure} jet target. The same normalization factor, 1.19, was then applied to the reported argon cross section, leaving room for further uncertainty. 
In addition,  it has to be pointed out that the results of RGF calculations, while describing both the LNF oxygen data~\cite{Meucci:2003} and the E12-14-012 argon data in the quasielastic region (see Fig.~\ref{xsec_Ar}), show the same normalization problem with the LNF argon data.

\begin{figure}[h!]
\centering
\includegraphics[width=1.0\columnwidth]{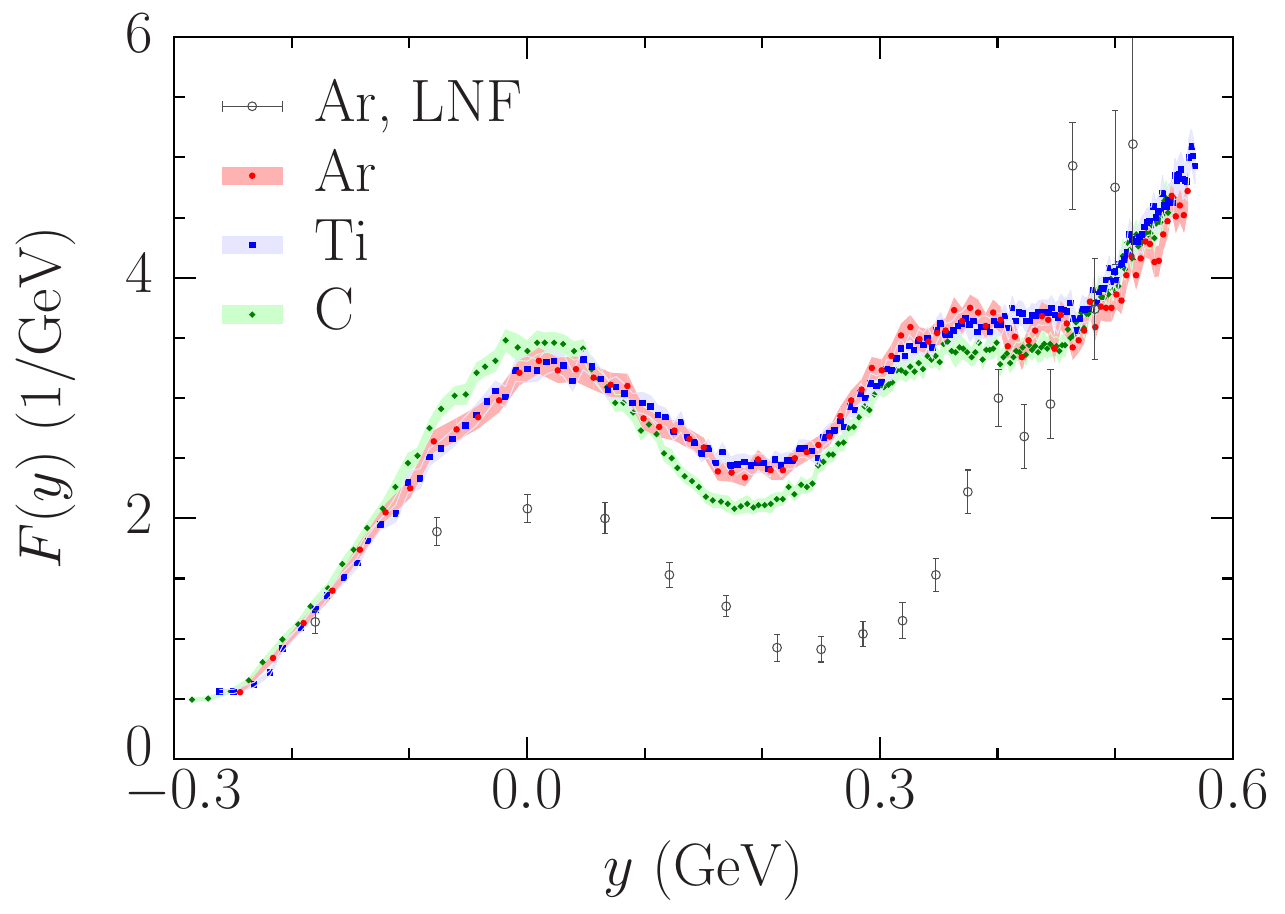}
\caption{(color online). Comparison between the scaling function $F(y)$ obtained from the E12-14-012 data on argon, titanium and carbon, and the argon data obtained at LNF~\cite{Anghinolfi:1995}.}
\label{fig:scaling}
\end{figure}

The pattern observed in Figs.\ref{fig:compare} and \ref{fig:superscaling} is also consistent with the results of Fig.~\ref{fig:scaling}, showing the 
scaling functions of the first kind, $F(y)$, obtained from the argon, titanium, and carbon cross section measured by the E12-14-012 collaboration, and from the argon cross section of Ref.\cite{Anghinolfi:1995}.  %
%}
%%%%%%%%%%%%%%%%%%%%%%%%%%%%%%%%%%%%%%%%%%%%%%%%%%%%%%%%%%%%%%%%%%%%%%%%

%
%-----------------------------------------------------------------------------------------------------------------------------------------------------------------------%
%
\par In this paper, we have reported the first argon results of JLab experiment E12-14-012, as {Ar}$(e,e^\prime )$ cross sections at incident electron energy $E=$ 2.222 GeV and scattering angle $\theta = 15.54^\circ$. The cross section covers a broad range of energy transfer in which quasielastic scattering and resonance production are the dominant mechanisms of interaction. We presented a comparison of the {Ar}$(e,e^\prime )$ cross section with previously reported {Ti}$(e,e^\prime )$ and {C}$(e,e^\prime )$ cross sections of our experiment. The new precise measurement on the argon nucleus will be of great value for the development of realistic models of the electroweak response of neutron-rich nuclei, vital for the success of the current and next generation of neutrino oscillation studies employing liquid-argon based detectors.
%
%-----------------------------------------------------------------------------------------------------------------------------------------------------------------------%
%
\par We acknowledge outstanding support from the Jefferson Lab Hall A technical staff, target group and Accelerator Division. This experiment was made possible by Virginia Tech and the National Science Foundation under CAREER Grant No. PHY$-$1352106. This work was also supported by the DOE Office of Science, Office of Nuclear Physics, Contract No.~DE-AC05-06OR23177, under which Jefferson Science Associates, LLC operates JLab, DOE contracts No. DE-FG02-96ER40950 and No. DE-AC02-76SF00515.
%
%-----------------------------------------------------------------------------------------------------------------------------------------------------------------------%

\end{document}